\RequirePackage{fix-cm}
\documentclass[onecolumn]{svjour3}
\smartqed
\usepackage[utf8]{inputenc}
\usepackage[T1]{fontenc}
\usepackage[hidelinks]{hyperref}
\usepackage{graphicx}
\usepackage{bm}
\usepackage{mathbbol}
\usepackage{braket}
\usepackage{amssymb,amsmath,amsfonts}
\usepackage{cite}

\newcommand{\up}{\uparrow}
\newcommand{\down}{\downarrow}

\begin{document}

\title{Trojan Quantum Walks}

\author{Henrique S. Ghizoni  \and Edgard P. M. Amorim}

\institute{H. S. Ghizoni   \at
Departamento de F\'isica, Universidade do Estado de Santa Catarina, 89219-710, Joinville, SC, Brazil \\
                           \and
           E. P. M. Amorim \at
Departamento de F\'isica, Universidade do Estado de Santa Catarina, 89219-710, Joinville, SC, Brazil\\
                           \email{edgard.amorim@udesc.br}
}

\date{Received: date / Accepted: date}

\maketitle

\begin{abstract}
We investigate the transport properties and entanglement between spin and position of one-dimensional quantum walks starting from a qubit over position states following a delta-like (local state) and Gaussian (delocalized state) distributions. We find out that if the initial state is delocalized enough and a NOT gate reflects this state backwards, then the interference pattern extinguishes the position dispersion without preventing the propagation of the state. This effect allows the creation of a Trojan wave packet, a non-spreading and non-stationary double-peak quantum state.


\end{abstract}

\section{Introduction} \label{sec:1}

From the epic poem \textit{Iliad}, the term "Trojan" was used to name a group of asteroids which share the Jupiter's orbit around the Sun. The center of mass of the Trojan asteroids is steady relative to Jupiter, once they are trapped on stable Lagrange regions ($L_4$ and $L_5$) of this celestial mechanical system \cite{emery2015complex}. In the quantum-mechanical context, Trojan wave packets have non-spreading and non-stationary behavior and they have been observed as a localized Rydberg electron over a circular orbit with dispersion suppressed by external fields \cite{birula1994lagrange,wyker2012creating}.

Quantum walks are the quantum analogue of the classical random walks \cite{aharonov1993quantum}. The quantum walker is a spin-$1/2$ particle positioned over discrete positions in a one-dimensional lattice. The quantum walk state is composed by a tensor product between the internal (spin) and external (position) degrees of freedom. The time evolution operator is formed by a quantum coin and a conditional displacement operator. While the quantum coin acts over the qubit (coin state), by putting it on a new superposition of spin states, the displacement operator translates the spin up (down) state to its right (left) neighbor position \cite{kempe2003quantum}. The main features of quantum walks are the ballistic dispersion being quadratically superior over their classical counterparts and the creation of entanglement between the spin and position \cite{venegas2012quantum}. 

Quantum walks have been widely studied as quantum search algorithms \cite{shenvi2003quantum} and a way to perform universal computation \cite{childs2009universal,lovett2010universal}. They also have provided some insightful ideas in applied social sciences \cite{busemeyer2006quantum} or for explaining the efficiency of photosynthesis \cite{engel2007evidence}. Lately, they have been used to model the neutrino flavor oscillations \cite{dimolfetta2016quantum} and foster new teleportation protocols \cite{wang2017generalized}. Moreover, they can be physically implemented in some promising experimental platforms \cite{wang2013physical}. 

Here, we study one-dimensional quantum walks starting from two kinds of initial position states (local and Gaussian) with a NOT gate on a specific position, which promotes a chiral reflection of the state. Curiously, we find out that a large initial delocalization and this reflection, together create a Trojan-like quantum state regardless of the initial qubit.

\section{Position-dependent quantum walks} \label{sec:2}

The Hilbert space of quantum walks is $\mathcal{H}=\mathcal{H}_C\otimes\mathcal{H}_P$ where $\mathcal{H}_C$ (coin space) is a complex two-dimensional vector space spanned by the spin states $\left\{\ket{\up},\ket{\down}\right\}$ and $\mathcal{H}_P$ (position space) is an infinite-dimensional and countably vector space spanned by $\left\{\ket{j}\right\}$ with $j\in\mathbb{Z}$ being a discrete position over a one-dimensional regular lattice. The initial quantum walk state has an initial coin state or qubit,
\begin{equation}
\ket{\Psi_C}=\cos\left(\dfrac{\alpha}{2}\right)\ket{\up}+e^{i\beta}\sin\left(\dfrac{\alpha}{2}\right)\ket{\down},
\label{qubit}
\end{equation}
over position states,
\begin{align}
\ket{\Psi(0)}&=\sum_{j=-\infty}^{+\infty}\ket{\Psi_C}\otimes f(j)\ket{j}\nonumber\\
&=\sum_{j=-\infty}^{+\infty}\left[a(j,0)\ket{\up}+b(j,0)\ket{\down}\right]\otimes \ket{j},
\label{Psi0}
\end{align}
where the summation is over all integers subject to $\sum_j[|a(j,0)|^2+|b(j,0)|^2]=1$ as the condition of normalization. The initial spin up and down amplitudes are $a(j,0)=f(j)\cos(\alpha/2)$ and $b(j,0)=f(j)e^{i\beta}\sin(\alpha/2)$ 
in the Bloch representation with $\alpha \in[0,\pi]$ and $\beta \in[0,2\pi]$ \cite{nielsen2010quantum} pondered by a position distribution function $f(j)$. Two particular initial states $\ket{\Psi(0)}$ addressed here are a local state with $f(j)=\delta(j)$ in eq. \eqref{Psi0},
\begin{equation}
\ket{\Psi_L(0)}=\ket{\Psi_C}\otimes\ket{0},
\label{Psi_L}
\end{equation}
and a Gaussian state,
\begin{equation}
\ket{\Psi_G(0)}=\sum_{j=-\infty}^{+\infty}\ket{\Psi_C}\otimes\dfrac{e^{-j^2/(4\sigma_0^2)}}{(2\pi\sigma_0^2)^{\frac{1}{4}}}\ket{j},
\label{Psi_0_Gauss}
\end{equation}
such that $\sigma_0$ is the initial dispersion of the state. 

The discrete unitary time-evolution of a quantum walk state gives
\begin{equation}
\ket{\Psi(n)}=\mathcal{T}\prod_{t=1}^{n}U(j-r)\ket{\Psi(0)},
\label{time_evolution}
\end{equation}
and $\ket{\Psi(n)}$ is the quantum walk state after $n$ time steps, $\mathcal{T}$ means a time-ordered product with the time $t$ starting from $1$ up to $n$ in discrete unitary increments and $U(j-r)$ is the position-dependent time-evolution operator given by, 
\begin{equation}
U(j-r)=\sum_{j=-\infty}^{+\infty}S\left[C(j-r)\otimes\ket{j}\bra{j}\right],
\label{op_evolution}
\end{equation}
where
\begin{equation}
S=\ket{\up}\bra{\up}\otimes\ket{j+1}\bra{j}+\ket{\down}\bra{\down}\otimes\ket{j-1}\bra{j},
\label{S_operator}
\end{equation}
is the conditional displacement operator which displaces the spin up (down) from the site $j$ to the site $j+1$ ($j-1$), and 
\begin{equation}
\displaystyle
C(j-r) = \dfrac{1}{\sqrt{2}}
\begin{bmatrix}
1 & 1 \\
1 & -1
\end{bmatrix}
+\dfrac{\delta(j-r)}{\sqrt{2}}
\begin{bmatrix}
-1         & \sqrt{2}-1 \\
\sqrt{2}-1 & 1
\end{bmatrix}
\label{Quantum_coin}
\end{equation}
is the position-dependent quantum coin and it operates over the spin states generating a superposition of them. Note that for $j\neq r$, the quantum coin is a Hadamard coin which creates a fair superposition between spin up and down without relative phases between them, and for a specific position $j=r$, the quantum coin is a NOT gate ($\sigma_x$ Pauli matrix). This position-defect reflects the state \cite{li2013position} in the opposite direction changing it from the spin up (down) to down (up) state.

Let us consider just one time step evolution of the quantum walk state in order to present the following recurrence relations for the spin up and down amplitudes,
\begin{align}
a(j,t)&=\frac{a(j-1,t-1)}{\sqrt{2}}\left[1-\delta(j-1-r)\right]+\frac{b(j-1,t-1)}{\sqrt{2}}\left[1+\delta(j-1-r)(\sqrt{2}-1)\right],\\
b(j,t)&=\frac{a(j+1,t-1)}{\sqrt{2}}\left[1+\delta(j+1-r)(\sqrt{2}-1)\right]+\frac{b(j+1,t-1)}{\sqrt{2}}\left[-1+\delta(j+1-r)\right],
\end{align}
therefore, starting from the initial amplitudes $a(j,0)$ and $b(j,0)$ is possible to calculate these amplitudes over all the positions and time steps by means of an iterative numerical procedure.

Since the state $\ket{\Psi(t)}$ remains pure over time, the entanglement between spin and position can be evaluated by the von Neumann entropy,
\begin{equation}
S_E(\rho(t))=-Tr[\rho_C(t)\log_2\rho_C(t)], 
\label{SE}
\end{equation}
of the partially reduced state $\rho_C(t)=Tr_P[\rho(t)]$ \cite{bennett1996concentrating} such that $\rho(t)=\ket{\Psi(t)}\bra{\Psi(t)}$ and $Tr_P[\cdot]$ gives the trace over the positions. Therefore, we have
\begin{equation}
\rho_C(t)= 
\begin{bmatrix}
A(t)   & B(t) \\
B^*(t) & 1-A(t)
\end{bmatrix},
\label{rho_C}
\end{equation}
with $A(t)=\sum_j|a(j,t)|^2$, $1-A(t)=\sum_j|b(j,t)|^2$ and $B(t)=\sum_ja(j,t)b^*(j,t),$ and $B^*(t)$ is the complex conjugate of $B(t)$. By diagonalizing $\rho_C(t)$, we have
\begin{equation}
S_E(\rho(t))=-\lambda_+(t)\log_2\lambda_{+}(t)-\lambda_{-}(t)\log_2\lambda_{-}(t),
\end{equation}
with eigenvalues $\lambda_{\pm}=1/2{\pm}[1/4-A(t)(1-A(t))+|B(t)|^2]^{1/2}$. The entanglement $S_E(t)$ can vary from $0$ for separable states up to $1$ for maximal entanglement between spin and position.

\section{Results}\label{sec:3}

The first main characteristic of quantum walks is their ballistic transport, whose dispersion is given by $\sigma(t)\propto t$, with different slopes of dispersion for quantum walks starting from distinct initial states or time-evolving by distinct quantum coins. The probability distribution can be symmetrical or not, depending on the initial state \cite{kempe2003quantum,venegas2012quantum}. The second main characteristic is the entanglement between spin and position, which also has a strong dependence on the initial state and used quantum coin. The initial qubits which lead to the maximal entanglement necessarily also lead to a symmetrical probability distribution regardless of the initial position state \cite{orthey2018connecting}. At the same time, while there are only two particular initial qubits over one position (local state) which lead the quantum walk to asymptotically time-evolve to the maximal entanglement condition, when a quantum walk starts from a delocalized state, there are infinite available initial qubits to reach this condition \cite{orthey2017asymptotic}.

\begin{figure}[h!]
\center\includegraphics[width=0.5\linewidth]{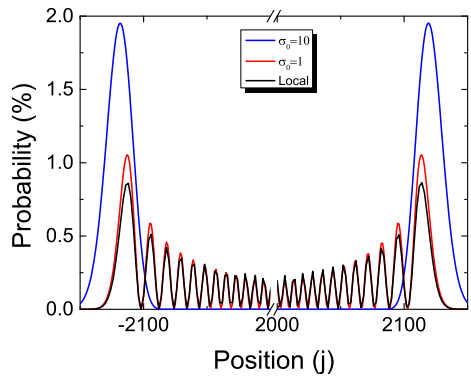}
\caption{Probability distribution over position for quantum walks starting from the qubit
$\ket{\Psi_C}=\cos(3\pi/8)\ket{\up}+\sin(3\pi/8)\ket{\down}$ with local (black) and Gaussian states with initial dispersion $\sigma_0=1$ (red) and $10$ (blue).}
\label{fig:1}
\end{figure}

Figure \ref{fig:1} shows the probability distribution for Hadamard quantum walks starting from local and two Gaussian states after $3000$ time steps. Numerically, we define the initial discrete Gaussian states between $j=-100$ and $100$, centered at $j=0$ and the condition of normalization for $\sigma_0=10$ gives a numerical error between $10^{-6}$ and $10^{-4}$ along the whole dynamics. The particular initial qubit employed here has $\alpha=3\pi/4$ and $\beta=0$ and it leads to a symmetrical probability distribution and also maximal entanglement for both position states \cite{orthey2017asymptotic,abal2006quantum,abal2006erratum}. One of the farthest peaks from the origin position has a corresponding probability of about $6\%$ for local and $15\%$ for the Gaussian state with $\sigma_0=1$, whereas close to $50\%$ with $\sigma_0=10$. Therefore, due to the initial delocalization of the state in this last case, almost the whole probability remains in only two peaks without appreciable probability amplitude between them. The distance over time between opposite peaks is about $\sqrt{2}t$ \cite{kempf2009group} in all cases here.

The vast majority of literature concerning quantum walks deals with a specific qubit placed on one position (local state) as initial state. Since the spreading and entanglement behavior of quantum walks have a strong dependence on the initial state, to evaluate their general features and trends, we study both position states (local and Gaussian) by averaging over $N$ initial qubits. Thus, in order to analyze the spreading behavior, the first quantity evaluated is the total average probability for each position $j$ after $t$ time steps, $\braket{|\Psi(j,t)|^2}=\sum_i (|a_i(j,t)|^2+|b_i(j,t)|^2)/N$ as a sum of the average probabilities of up and down states respectively, with the index $i$ varying for all initial qubits considered. The second average quantity is the dispersion, $\braket{\sigma(t)}=\sum_i\sigma_i(t)/N$, and, in the same way, the average entanglement is given by $\braket{S_E(t)}=\sum_iS_E^i(t)/N$. 

We perform calculations of quantum walks with $3000$ time steps starting from a local state and Gaussian states with initial dispersion $\sigma=1$ and $10$. The coins we used here are Hadamard and $C(j+101)$, i.e., a NOT gate as a quantum coin placed on $j=-101$ which performs a chiral reflection of the state and also a Hadamard coin for all positions $j\geq-100$. Our results were obtained by averaging over a set of $N=2,016$ qubits starting from $(\alpha,\beta)=(0,0)$ to $(\pi,2\pi)$ with independent increments of $0.1$. This is the same procedure used in earlier works \cite{orthey2017asymptotic,orthey2018weak,vieira2013dynamically,vieira2014entangling}. 

\begin{figure*}[h!]
\center\includegraphics[width=\linewidth]{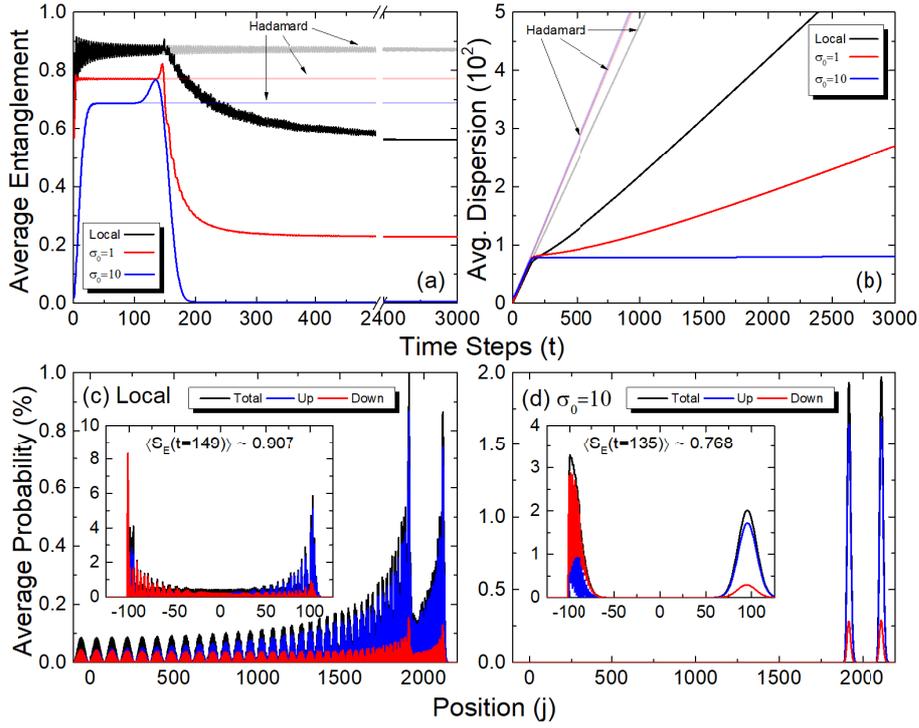}
\caption{Average (a) entanglement $\braket{S_E(t)}$ and (b) dispersion $\braket{\sigma(t)}$ over time for quantum walks starting from local (black) and Gaussian states with initial dispersion $\sigma_0=1$ (red) and $10$ (blue). These states evolve by means of a Hadamard coin (transparent lines) and $C(j+101)$ (solid lines), which corresponds to a Hadamard coin over all positions except for a NOT gate in $j=-101$. In (a) there is a break-region within $t=500$ and $2400$. Average probability distribution for (c) local state and (d) Gaussian state with $\sigma_0=10$ showing total (black), spin up (blue) and down (red) probabilities. The insets in (c) and (d) show the average probability for the corresponding time step of the maximum entanglement situation for each case.}
\label{fig:2}
\end{figure*}

\begin{figure}[h!]
\center\includegraphics[width=0.5\linewidth]{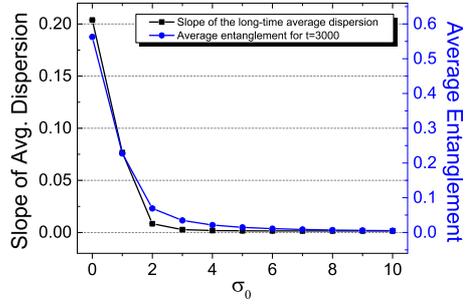}
\caption{Slope of the long-time average dispersion regime $\braket{\sigma(t)}$ (black square) obtained from a linear fitting made from the last $2000$ time steps, and average entanglement (blue circle) $\braket{S_E(t=3000)}$ over the initial dispersion of the state. The points where $\sigma_0=0$ correspond to the local state. The lines connecting points are just a guide for the eyes.}
\label{fig:3}
\end{figure}

Figure \ref{fig:2} shows (a) the average entanglement and (b) dispersion over time for all cases. For Hadamard walks, when the states achieve the long-time entanglement regime, the entanglement values are of about $0.87$, $0.77$ and $0.69$ (transparent lines). After the states are reflected in $j=-101$ (solid lines), they reach maximum entanglement values of about $0.91$, $0.82$ and $0.77$ when approximately one quarter of the states interfere with themselves, as can be seen in the insets in (c) and (d), followed by a substantial entanglement drop to values of about $0.57$ for local state, $0.23$ and smaller than $0.01$ for Gaussian states with $\sigma_0=1$ and $10$ respectively. The long-time dispersion behavior is always linear, but the reflection of the states reveals a significant change in the slope of the dispersion for all cases, since the relative velocity between opposite peaks vanishes and a notable positive group velocity of the state appears. However, at this point, we have an important difference between the local and Gaussian states: while the (c) local state continues to spread, the (d) Gaussian state with large initial delocalization ($\sigma_0=10$) prevents the dispersion of the state allowing the creation of a Trojan wave packet at the cost of the loss of the quantum correlation. Figure \ref{fig:3} makes a comparison between the average slope of the long-time dispersion and the entanglement after $3000$ time steps over the initial dispersion $\sigma_0$ of the state.

\section{Conclusion}\label{sec:4}

In summary, we have performed some numerical calculations which show that quantum walks can also exhibit a non-spreading and non-stationary behavior. When quantum walks start from a Gaussian state with a large enough initial dispersion, they time-evolve with only two opposite peaks without considerable amplitudes of probability between them. After one of the peaks is reflected by a NOT gate on a particular position, the relative velocity between peaks vanishes creating a double-peak Trojan wave packet without quantum correlation (entanglement) between internal and external degrees of freedom. At last, we hope our findings can be used to foster the discussion about the creation of coherent states, the relation between entanglement and coherence \cite{streltsov2015measuring}, the quantum-classical limits of such walks and the experimental researchers can corroborate our results.

\section*{Acknowledgements}
This study was financed in part by the Coordenação de Aperfeiçoamento de Pessoal de Nível Superior - Brasil (CAPES) - Finance Code 001. HSG and EPMA thank Janice Longo for her careful reading of the manuscript.

\end{document}